\documentclass[epjCONF]{svjour}
\usepackage{graphicx}
\usepackage[varg]{txfonts}
\usepackage{bm}
\usepackage[latin1]{inputenc}
\session-title{%
19$^{\textnormal{\footnotesize th}}$ International %
IUPAP Conference on Few-Body Problems in Physics%
}
\begin{document}
\title{%
Electromagnetic Structure and Reactions of Few-Nucleon Systems in $\chi$EFT
}%
\author{%
L.\ Girlanda\inst{1,2} 
\and %
S.\ Pastore\inst{3} 
\and %
R.\ Schiavilla\inst{3,4}\fnmsep\thanks{\email{schiavil@jlab.org}} 
\and %
M.\ Viviani\inst{2}}
\institute{%
Department of Physics, University of Pisa, 56127 Pisa, Italy
\and %
INFN-Pisa, 56127 Pisa, Italy
\and %
Department of Physics, Old Dominion University, Norfolk, VA 23529, USA
\and %
Theory Center, Jefferson Lab, Newport News, VA 23606, USA
}
\abstract{
We summarize our recent work dealing with the construction of the nucleon-nucleon
potential and associated electromagnetic currents up to one loop in chiral effective
field theory ($\chi$EFT).  The magnetic dipole operators derived from these currents
are then used in hybrid calculations of static properties and low-energy
radiative capture processes in few-body nuclei.  A preliminary set of results
are presented for the magnetic moments of the deuteron and trinucleons
and thermal neutron captures on $p$, $d$, and $^3$He.
} 
\maketitle
\section{Introduction}
\label{sec:intro}

The non-perturbative character of quantum chromodynamics (QCD)
at low energies has so far prevented a quantitative understanding
of nuclear structure and reactions in terms of the theory fundamental degrees
of freedom, {\it i.e.} quarks and gluons.  However, the
chiral symmetry exhibited by QCD severely restricts the form of
the interactions of pions among themselves and with other
particles~\cite{Weinberg95}.  In particular, the pion couples to
the baryons, such as nucleons or $\Delta$-isobars, by powers of its
momentum $Q$, and the Lagrangian describing these interactions
can be expanded in powers of $Q/\Lambda_\chi$, where $\Lambda_\chi \sim 1$
GeV specifies the chiral-symmetry breaking scale.  As a consequence,
classes of Lagrangians emerge, each characterized by a given power
of $Q/\Lambda_\chi$ and each involving a certain number of
unknown coefficients, so called low-energy constants (LEC's), which are then
determined by fits to experimental data (see, for example, the
review papers~\cite{Epelbaum08}, and
references therein).

This approach, known as chiral effective field theory ($\chi$EFT),
has been used to study two- and many-nucleon interactions~\cite{Epelbaum08}
and the interaction of electroweak probes with nuclei~\cite{Park93,Park96}.
Its validity, though, is restricted to processes occurring at low energies.
In this sense, it has a more limited range of applicability than meson-exchange
or more phenomenological models of these interactions, which in fact
quantitatively and successfully account for a wide variety of nuclear
properties and reactions up to energies, in some cases, well beyond
the pion production threshold (for a review, see Ref.~\cite{Carlson98}).
However, it can be justifiably argued that
$\chi$EFT puts nuclear physics on a more fundamental basis
by providing, on the one hand, a direct connection between
QCD and its symmetries, in particular chiral symmetry, and the
strong and electroweak interactions in nuclei, and, on the
other hand, a practical calculational scheme susceptible,
in principle, of systematic improvement.

The present report summarizes recent work carried out by our group
in the construction of the nucleon-nucleon ($NN$) potential and associated electromagnetic
currents up to one loop in $\chi$EFT~\cite{Pastore08,Pastore09}, and in their
application to the calculation of magnetic dipole ($M1$) observables in
$A$=2--4 nuclei~\cite{Girlanda09}.  The derivation of the potential and
currents is based on time-ordered perturbation theory and the non-relativistic
Hamiltonians implied by the chiral Lagrangians of Refs.~\cite{Weinberg90,vanKolck94,Epelbaum98},
and retains irreducible as well as recoil-corrected reducible diagrams.
The latter arise from expanding the energy denominators
$( \Delta E_N+\omega_\pi)^{-1}$, where $\Delta E_N$ and
$\omega_\pi$ denote, respectively, nucleon kinetic energy differences and pion energies,
in powers of $\Delta E_N/\omega_\pi$ (which is of order $Q$, the low-momentum scale).
We will not discuss some of the more technical aspects of the formalism, including, for example,
the renormalization of loop corrections in dimensional regularization or
the cancellations occurring between irreducible and recoil-corrected reducible contributions.
Some of these issues were outlined in the talk~\cite{Schiavilla09}, but
the interested reader may want to consult the original papers~\cite{Pastore08,Pastore09}.

We will also not attempt to cite all of the extensive literature on these topics---most of which,
incidentally, can be found in the review papers mentioned above---but rather will refer
only to those papers we are familiar with and which are directly relevant to our work.

\section{$NN$ potential at one loop}

\begin{figure*}[!htb]
\centering
\includegraphics[width=1.25\columnwidth]{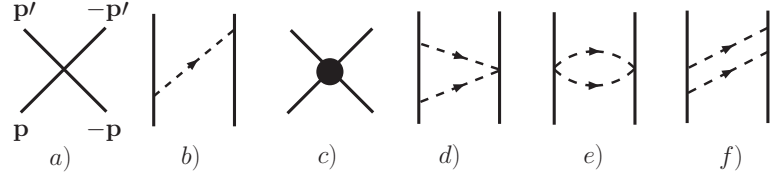}
\caption{Diagrams illustrating contributions to the $NN$ potential
entering at LO ($Q^{\,0}$), panels a) and b), and N$^2$LO ($Q^{\,2}$), panels c)-f).
Nucleons and pions are denoted by solid and dashed lines,
respectively.  The filled circle in panel c) represents the vertex from
contact Hamiltonians containing two gradients of the nucleons' fields.
Only one among the possible time orderings is shown for
each contribution with more than one vertex.}
\label{fig:fig1}
\end{figure*}

In Fig.~\ref{fig:fig1} we show the diagrams illustrating the contributions
occurring up to N$^2$LO.  At LO ($Q^0$) there is a contact interaction,
panel a), along with the one-pion-exchange contribution, panel b).
At N$^2$LO ($Q^2$) there are i) contact interactions involving two gradients acting on the nucleons'
fields, panel c), and ii) two-pion-exchange loop contributions,
panels d)-f).
After renormalization, we find that the potential in the center-of-mass frame
is given by~\cite{Pastore09,vanKolck94,Epelbaum98}
\begin{equation}
v({\bf k},{\bf K})={v}^{\rm CT0}+{v}^\pi({\bf k})+{v}^{\rm CT2}({\bf k},{\bf K})
+{v}^{2\pi}({\bf k}) 
\label{eq:vn2lo}
\end{equation}
where
\begin{equation}
\label{eq:ct0}
v^{\rm CT0}= C_S+C_T\,{\bm \sigma}_1\cdot{\bm \sigma}_2 
\end{equation}
\begin{equation}
\label{eq:ope}
v^{\pi}({\bf k})=-\frac{g_A^2}{F_{\pi}^2}\,{\bm \tau}_1\cdot{\bm \tau}_2\,
\frac{{\bm \sigma}_1\cdot{\bf k}\,{\bm\sigma}_2\cdot{\bf k}}{m_\pi^2+k^2} 
\end{equation}
\begin{eqnarray}
v^{\rm CT2}({\bf k}, {\bf K})&=& C_1\,k^2+C_2\,K^2+
(C_3\,k^2+C_4\,K^2)\,{\bm \sigma}_1\cdot{\bm \sigma}_2  \nonumber\\
&+& i\,C_5\,\frac{{\bm \sigma}_1+{\bm \sigma}_2}{2}\cdot {\bf K}\times{\bf k}
+C_6\,{\bm \sigma}_1\cdot{\bf k}\,\,{\bm \sigma}_2\cdot {\bf k} \nonumber \\
&+&C_7\,{\bm \sigma}_1\cdot{\bf K}\,\,{\bm \sigma}_2\cdot {\bf K} 
\label{eq:ct2} 
\end{eqnarray}
\begin{eqnarray}
{v}^{2\pi}({\bf k})&=&\frac{1}
{48 \pi^2\,F_{\pi}^4}{\bm \tau}_1\cdot{\bm \tau}_2\,
G(k) \, \Bigg[4\, m_{\pi}^2(1+4\, g_A^2-5\, g_A^4) \nonumber \\
&+&k^2(1+10\, g_A^2 - 23\, g_A^4)-\frac{48\,g_A^4 m^4_\pi}
{4\,  m^2_\pi+k^2}\Bigg] \nonumber \\
&+&\frac{3\,g_A^4}{8\pi^2\,F_{\pi}^4}\,G(k)\, \left( k^2\,{\bm \sigma}_1\cdot{\bm \sigma}_2-
{\bm \sigma}_1 \cdot{\bf k}\, {\bm \sigma}_2\cdot{\bf k}\right) \ ,
\label{eq:tper}
\end{eqnarray}
the momenta ${\bf k}$ and ${\bf K}$ are defined
in terms of the nucleons' initial and final relative momenta
${\bf p}$ and ${\bf p}^{\prime}$ as
${\bf k} = {\bf p}^{\prime} -{\bf p}$ and
${\bf K}=({\bf p}^{\prime}+{\bf p})/2$,
$g_A$ and $F_\pi$ are the nucleon axial coupling constant and pion decay amplitude,
respectively, and $C_S$, $C_T$, and $C_i$ are LEC's.
The function $G(k)$ reads
\begin{equation}
G(k)=\frac{\sqrt{4\,m_{\pi}^2+k^2}}{k}\ln \frac{\sqrt{4\,m_{\pi}^2+k^2}+k}{\sqrt{4\,m_{\pi}^2+k^2}-k} \ .
\label{eq:loopf}
\end{equation}

Before turning our attention to a discussion of the phase shifts, we note
that the potential above needs to be regularized
because of its power-law behavior for large values of the momenta $k$ and/or
$K$.  This is accomplished by including a high-momentum cutoff, which
we take to be of the form
\begin{equation}
 C_\Lambda(k,K)={\rm e}^{-(k^4+16\,K^4)/\Lambda^4} 
\end{equation}
so that the matrix elements of the regularized potential
entering the $K$-matrix and bound-state equations are
obtained from
\begin{equation}
v^{\rm R}({\bf k},{\bf K})=v({\bf k},{\bf K})\,C_\Lambda(k,K) \ .
\end{equation}
In the following
cutoff parameters $\Lambda$ in the range 500--700 MeV are considered.
Thus $C_\Lambda(k,K)$ removes momenta larger than (3--$4)\, m_\pi$ in
a theory retaining up to two-pion-exchange mechanisms, and whose regime
of validity extends, therefore, up to $2\, m_\pi$.

The LEC's $C_S$, $C_T$, and $C_i$ are determined by fitting
the deuteron binding energy and  S- and P-wave $np$ phase shifts up to
laboratory kinetic energies of 100 MeV, as obtained in the very recent (2008)
analysis of Gross and Stadler~\cite{Gross08}.
\begin{table}
\caption{Values for the nucleon axial coupling constant $g_A$, pion
decay constant $F_\pi$,  neutral and charged pion masses $m_0$ and
$m_+$, and (twice) $np$ reduced mass $\mu_N$, used in the fits.}
\label{tb:cons}
\begin{tabular}{lcccc}
\hline\noalign{\smallskip}
$g_A$  & $F_\pi$ (MeV)  & $m_0$ (MeV)  &  $m_+$ (MeV)  & $2\, \mu_N$ (MeV)  \\
\noalign{\smallskip}\hline\noalign{\smallskip}
   1.29    &    184.8    &  134.9766      &  139.5702      & 938.9181    \\
\noalign{\smallskip}\hline
\end{tabular}
\end{table}
The parameters characterizing the one- and
two-pion exchange parts of the potential are listed in Table~\ref{tb:cons}, with
$g_A$ determined from the Golberger-Treiman
relation $g_A=g_{\pi NN}F_\pi/(2\, m_N)$, where the $\pi NN$ coupling constant
is taken to have the value $g^2_{\pi NN}/(4\pi)=13.63 \pm 0.20$~\cite{Stoks93a,Arndt94}.
In fact, in the one-pion exchange we include the isospin-symmetry breaking induced by the
mass difference between charged and neutral pions, since it leads to significant
effects in the $^1$S$_0$ scattering length~\cite{Wiringa95}, and therefore the one-pion-exchange
potential reads
\begin{eqnarray}
{v}^{\pi}({\bf k})&=&-\frac{g_A^2}{3\, F_{\pi}^2} \left[ {\bm \tau}_1\cdot{\bm \tau}_2
\Bigg( \frac{1}{k^2+m_0^2}+\frac{2}{k^2+m_+^2} \right) \nonumber \\
&+&T_{12} \left( \frac{1}{k^2+m_0^2}-\frac{1}{k^2+m_+^2} \Bigg)
 \right] {\bm \sigma}_1\cdot{\bf k}\,{\bm\sigma}_2\cdot{\bf k}
\end{eqnarray}
where $T_{12}$ is the isotensor operator defined as
\begin{equation}
T_{12}=3\, \tau_{1,z}\tau_{2,z}-{\bm \tau}_1 \cdot {\bm \tau}_2
\end{equation}
and $m_0$ and $m_+$ are the neutral and charged
pions masses.  Finally, we note that the pion mass entering in the two-pion-exchange part
is taken as $m_\pi=( m_0+2\, m_+)/3$.

\begin{table}
\caption{Values of the LEC's corresponding to cutoff parameters $\Lambda$ in the range 500--700 MeV,
obtained from fits to $np$ phase shifts up to lab energies of 100 MeV.}
\label{tb:lec}
\begin{tabular}{lccc}
\hline\noalign{\smallskip}
& \multicolumn{3}{c}{$\Lambda$ (MeV)} \\
\noalign{\smallskip}\hline
& \multicolumn{1}{c}{500} & \multicolumn{1}{c}{600}  & \multicolumn{1}{c}{700} \\
\hline
$C_S$ (fm$^2$)  &   --4.456420   &   --4.357712   &   --3.863625   \\
$C_T$ (fm$^2$)  &    0.034780   &    0.094149   &    0.234176   \\
$C_1$ (fm$^4$)  &   --0.360939   &   --0.259186   &   --0.268296   \\
$C_2$ (fm$^4$)  &   --1.460509   &   --0.934505   &   --0.835226   \\
$C_3$ (fm$^4$)  &   --0.349780   &   --0.359547   &   --0.389047   \\
$C_4$ (fm$^4$)  &   --1.968636   &   --1.717178   &   --1.724544   \\
$C_5$ (fm$^4$)  &   --0.870067   &   --0.754021   &   --0.695564   \\
$C_6$ (fm$^4$)  &    0.326169   &    0.301194   &    0.348152   \\
$C_7$ (fm$^4$)  &  --0.727797   &   --1.006459   &   --0.955273   \\
\noalign{\smallskip}\hline
\end{tabular}
\end{table}

\begin{figure*}[bhtb]
\centering
\includegraphics[width=1\columnwidth]{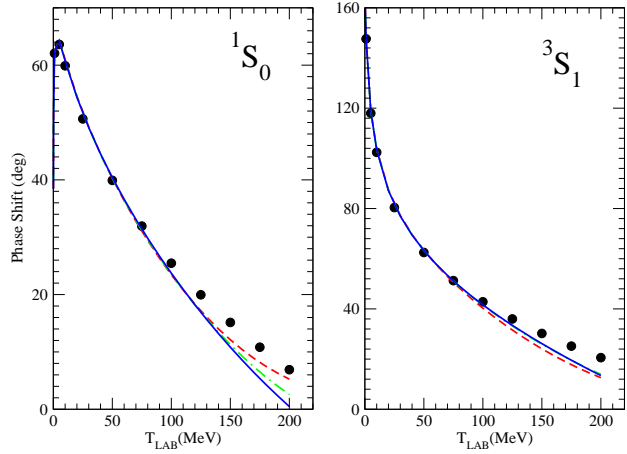}
\vspace{0.25cm}
\caption{The S-wave $np$ phase shifts, obtained with cutoff parameters
$\Lambda$=500, 600, and 700 MeV, are denoted by dash (red),
dot-dash (green), and solid (blue) lines, respectively.
The filled circles represent the phase-shift analysis of Ref.~\protect\cite{Gross08}.}
\label{fig:sw}
\end{figure*}

\begin{figure*}[bthp]
\centering
\includegraphics[width=1.5\columnwidth]{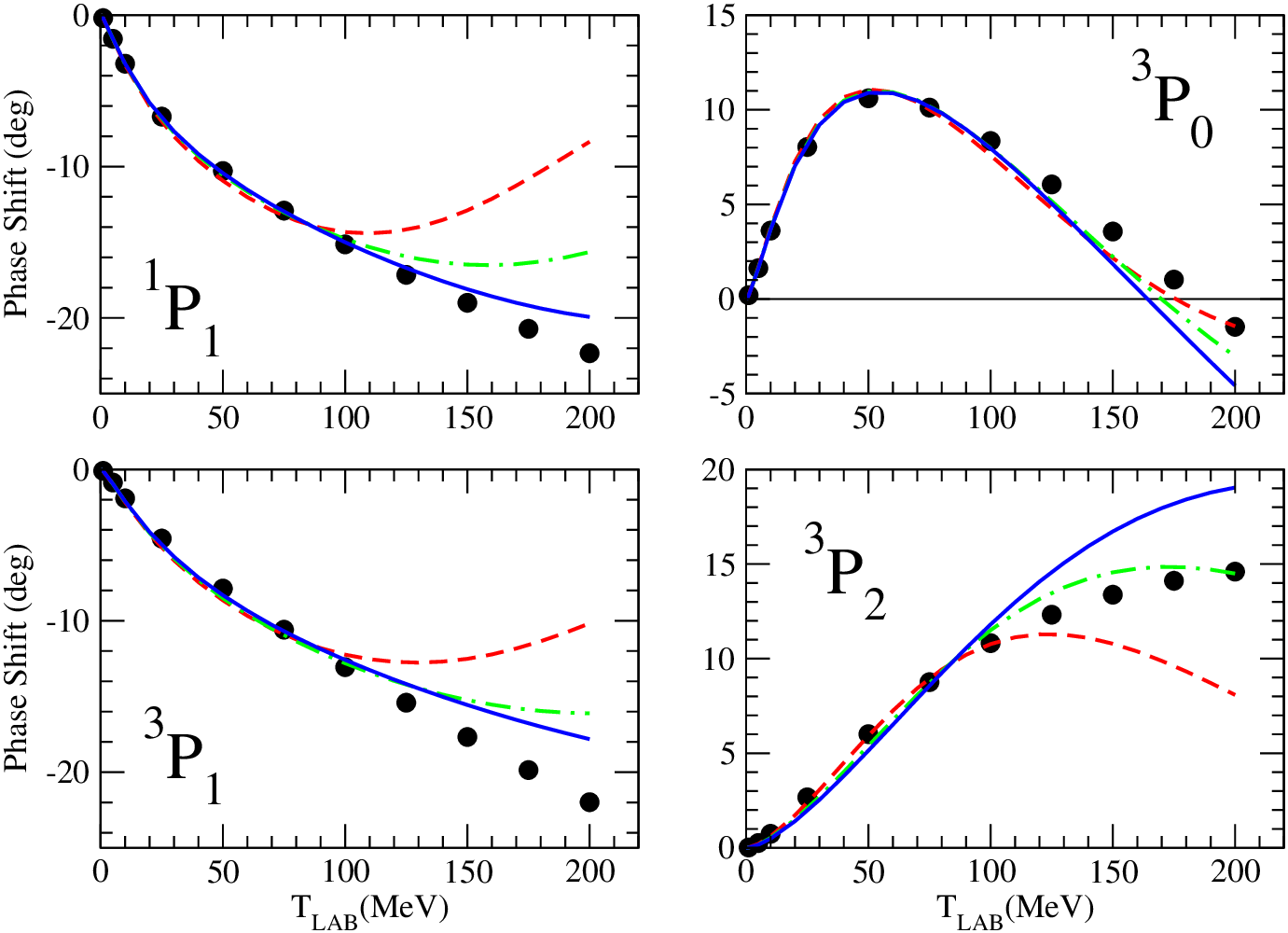}
\vspace{0.25cm}
\caption{Same as in Fig.~\protect\ref{fig:sw}, but for P-wave phase shifts.}
\label{fig:pw}
\vspace{1.7cm}
\includegraphics[width=1.5\columnwidth]{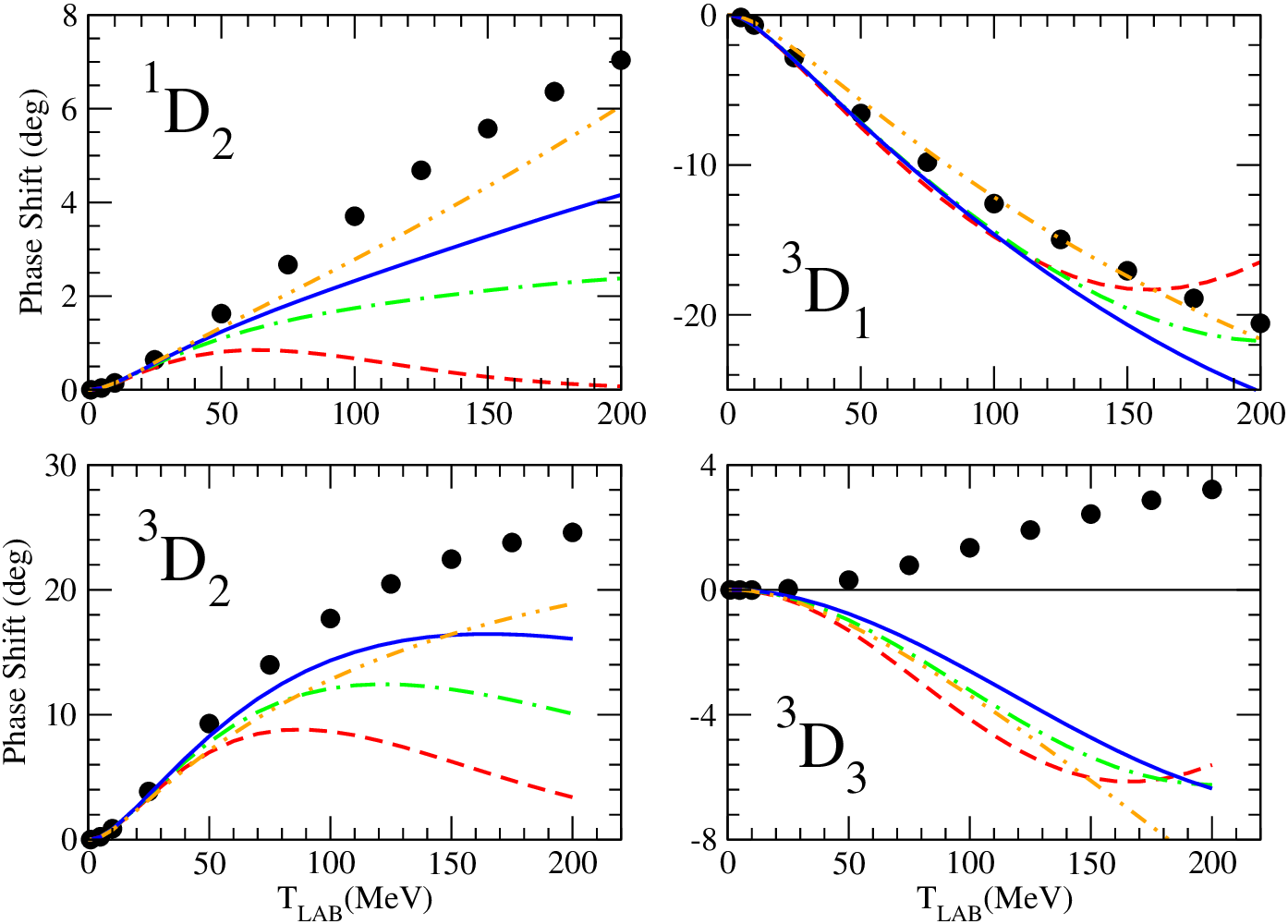}
\vspace{0.25cm}
\caption{Same as in Fig.~\protect\ref{fig:sw}, but for D-wave phase shifts.
The dash-double-dot (orange) line
is obtained in first order perturbation theory for the $T$-matrix by including only the
one- and two-pion-exchange parts of the N$^2$LO potential. }
\label{fig:dw}
\end{figure*}

\begin{figure*}[bthp]
\centering
\includegraphics[width=1.5\columnwidth]{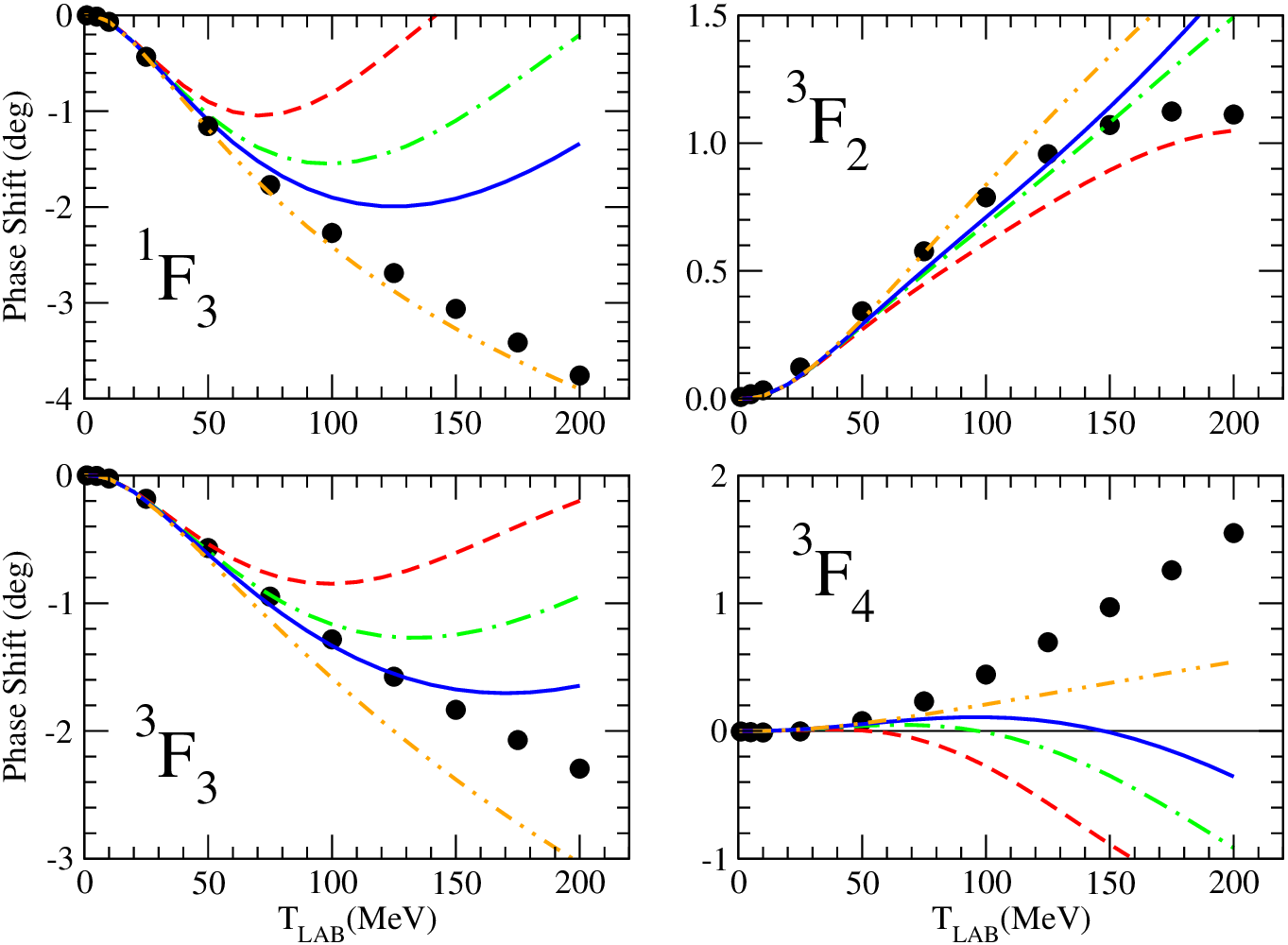}
\vspace{0.25cm}
\caption{Same as in Fig.~\protect\ref{fig:dw}, but for F-wave phase shifts. }
\label{fig:fw}
\vspace{1.7cm}
\includegraphics[width=1.5\columnwidth]{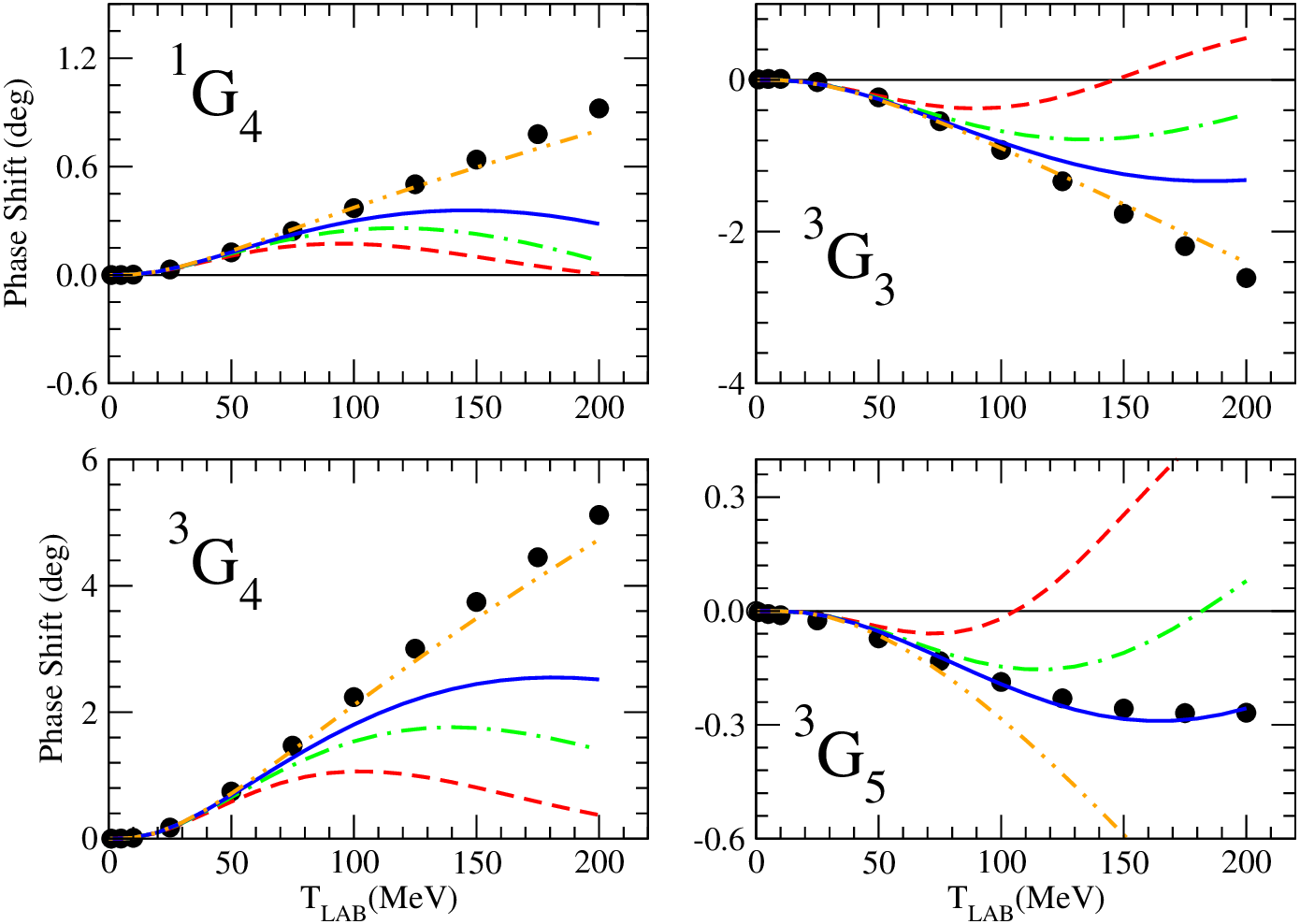}
\vspace{0.25cm}
\caption{Same as in Fig.~\protect\ref{fig:dw}, but for G-wave phase shifts.}
\label{fig:gw}
\end{figure*}

\begin{figure*}[bthp]
\centering
\includegraphics[width=1.5\columnwidth]{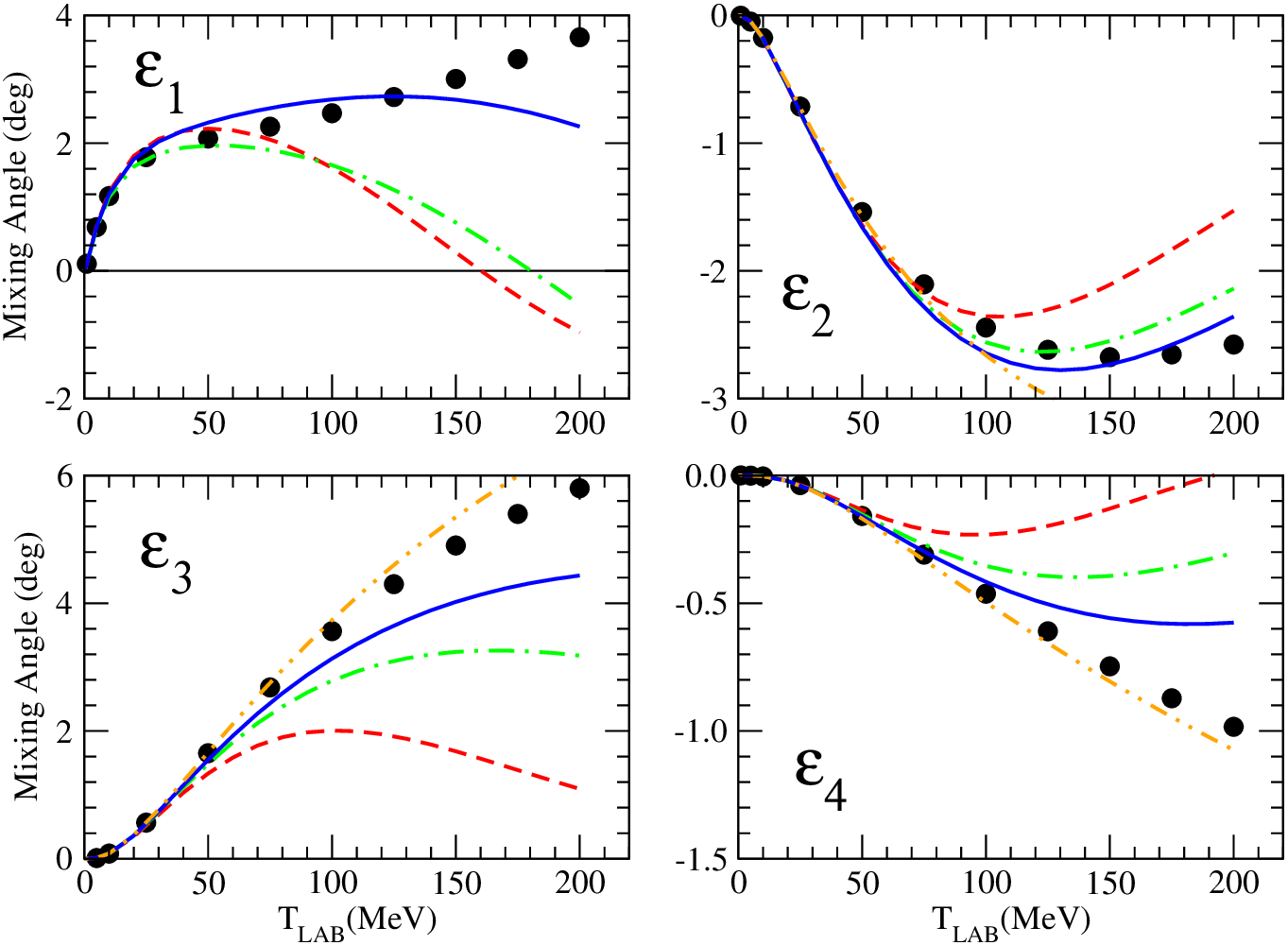}
\vspace{0.25cm}
\caption{Same as in Fig.~\protect\ref{fig:dw}, but for the mixing angles $\epsilon_J$.}
\label{fig:ew}
\vspace{1.7cm}
\includegraphics[width=1.5\columnwidth]{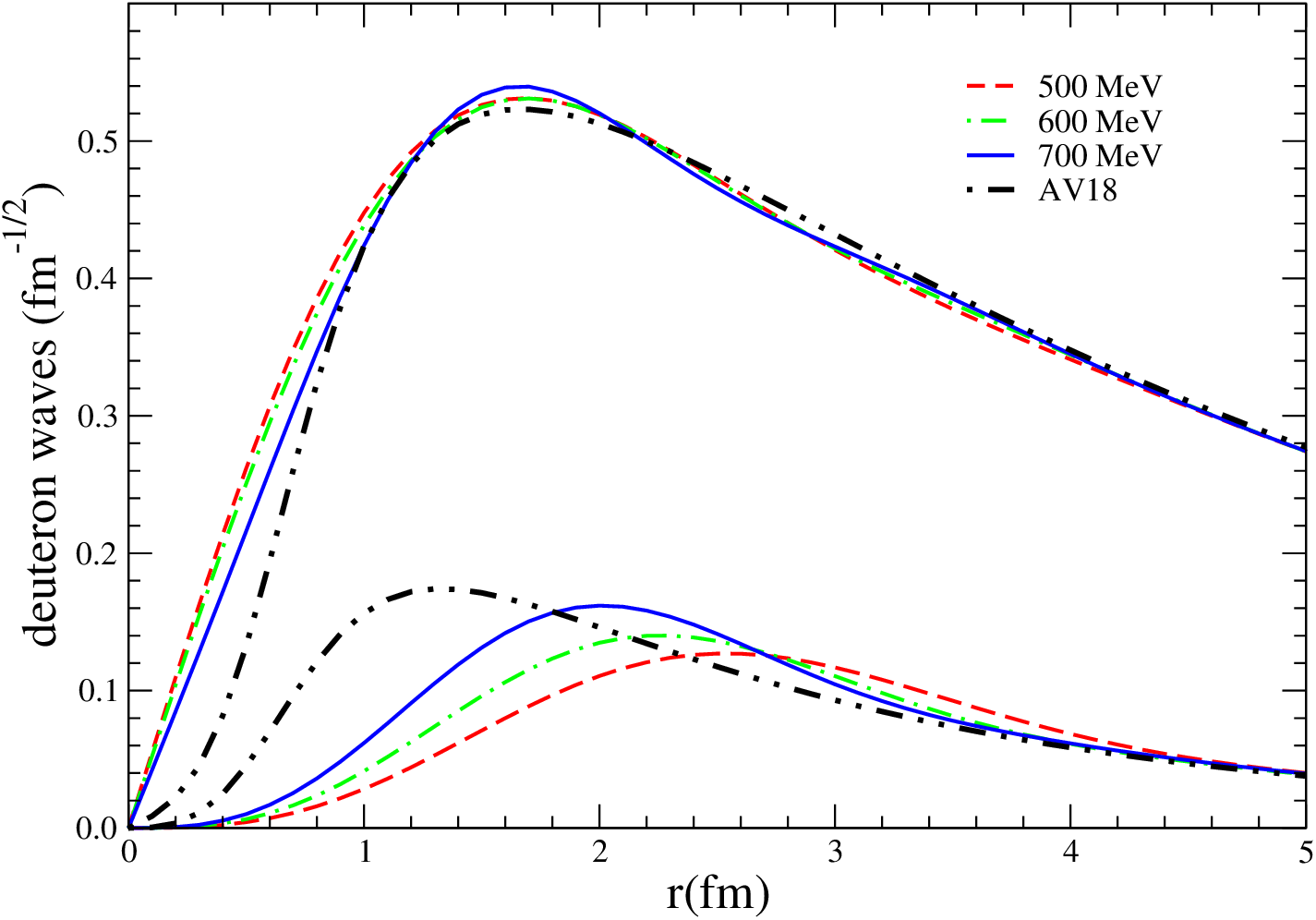}
\vspace{0.25cm}
\caption{(Color online) The S-wave and D-wave components
of the deuteron, obtained with cutoff parameters $\Lambda$=500, 600, and 700 MeV
and denoted by dash (red), dot-dash (green), and solid (blue) lines,
respectively, are compared with those calculated
from the Argonne $v_{18}$ potential (dash-double-dot black lines).}
\label{fig:deut}
\end{figure*}

The best-fit values obtained for the LEC's are listed in Table~\ref{tb:lec} for $\Lambda$=500, 600,
and 700 MeV, while results for the S- and P-wave phases
used in the fits, as well as for the D-wave and
peripheral  F- and G-wave phases, and mixing angles $\epsilon_{J=1,\dots,4}$
are displayed in Figs.~\ref{fig:sw}--\ref{fig:ew} up to 200 MeV lab kinetic energies.
Effective range expansions and deuteron properties are listed in Table~\ref{tb:erd}.
For reference, in Figs.~\ref{fig:dw}--\ref{fig:ew}, following the
original work by Kaiser {\it et al.}~\cite{Kaiser97}, the phases obtained by including
only the one- and two-pion-exchange (${v}^\pi$ and
${v}^{2\pi}$, respectively) terms of the potential are also shown.
These have been calculated in first order perturbation
theory on the $T$-matrix, and hence are cutoff independent.
Overall, the quality of the fits at N$^2$LO
is comparable to that reported in Refs.~\cite{Epelbaum00,Entem02} and,
more recently, in Ref.~\cite{Yang09}.
While the cutoff dependence is relatively weak for the S-wave phases
beyond lab energies of 100 MeV, it becomes significant for higher
partial wave phases and for the mixing angles.  In particular, the F- and G-wave phases,
while small because of the centrifugal barrier, nevertheless display a
pronounced sensitivity to short-range physics, although there are indications~\cite{Kaiser98}
that inclusion of explicit $\Delta$-isobar degrees of freedom might
reduce this sensitivity.
Beyond 100 MeV, the agreement between the calculated and experimental phases
is generally poor, and indeed in the $^3$D$_3$ and $^3$F$_4$ channels
they have opposite sign.  The scattering lengths are well reproduced
by the fits (within $\sim 1$\% of the data, see Table~\ref{tb:erd}),
however, the singlet and triplet effective ranges are both significantly
underpredicted, by $\sim 10$\% and $\sim 5$\% respectively.
\begin{table}
\caption{Singlet and triplet $np$ scattering lengths ($a_s$ and $a_t$) and effective
ranges ($r_s$ and $r_t$), and deuteron binding energy ($B_d$), D- to S-state ratio
($\eta_d$), root-mean-square matter radius ($r_d$), magnetic moment ($\mu_d$), quadrupole
moment ($Q_d$), and D-state probability ($P_D$), obtained
with $\Lambda$=500, 600, and 700 MeV, are compared to the corresponding experimental
values.}
\label{tb:erd}
\begin{tabular*}{0.48\textwidth}{lcccc}
\hline\noalign{\smallskip}
& \multicolumn{3}{c}{$\Lambda$ (MeV)} \\
\hline\noalign{\smallskip}
& \multicolumn{1}{c}{500} & \multicolumn{1}{c}{600}  & \multicolumn{1}{c}{700}
& \multicolumn{1}{c}{Expt} \\
\noalign{\smallskip}\hline\noalign{\smallskip}
$\!\!\!a_s$ (fm)         &   --23.729   &   --23.736   &   --23.736   &   --23.749(8)   \\
$\!\!\!r_s$ (fm)         &     2.528   &     2.558   &     2.567   &     2.81(5)    \\
$\!\!\!a_t$ (fm)         &     5.360   &     5.371   &     5.376   &     5.424(3)   \\
$\!\!\!r_t$ (fm)         &     1.665   &     1.680   &     1.687   &     1.760(5)   \\
$\!\!\!B_d$ (MeV)        &     2.2244  &     2.2246  &     2.2245  &     2.224575(9)\\
$\!\!\!\eta_d$           &     0.0267  &     0.0260  &     0.0264  &     0.0256(4)  \\
$\!\!\!r_d$ (fm)         &     1.943   &     1.947   &     1.951   &     1.9734(44) \\
$\!\!\!\mu_d$ ($\mu_N$)  &     0.860   &     0.858   &     0.853   &     0.8574382329(92)\\
$\!\!\!Q_d$ (fm$^2$)     &     0.275   &     0.272   &     0.279   &     0.2859(3)  \\
$\!\!\!P_D$ (\%)         &     3.44    &     3.87    &     4.77    &                \\
\noalign{\smallskip}\hline
\end{tabular*}
\end{table}

The deuteron S- and D-wave radial wave functions are shown in Fig.~\ref{fig:deut}
along with those calculated with the Argonne $v_{18}$ (AV18) potential~\cite{Wiringa95}.
The D wave is particularly sensitive to variations in the
cutoff: it is pushed in as $\Lambda$ is increased from 500 to 700 MeV,
but remains considerably smaller than that of the AV18 up to internucleon
distances of $\sim 1.5$ fm, perhaps not surprisingly, since this realistic
potential has a strong tensor component at short range.  The static
properties, {\it i.e.} D- to S-state ratio, mean-square-root matter
radius, and magnetic moment (the binding energy is fitted) are
close to the experimental values, and their variation with $\Lambda$
is quite modest.  The quadrupole moment is underpredicted by $\sim 4$\%,
a pathology common, to the best of our knowledge, to all realistic potentials
(including the AV18).

\section{Magnetic moments at one loop}
\label{sec:cnts}

The LO term in the electromagnetic current operator
results from the coupling of the external photon field to the individual
nucleons, and is counted as $e\,Q^{-2}$ ($e$ is the electric charge),
where a factor $e\,Q$ is from the $\gamma NN$ vertex,
and a factor $Q^{-3}$ follows from the momentum $\delta$-function implicit in this type
of disconnected diagrams.  It consists of the standard convection and spin-magnetization
currents of the nucleon.  The NLO term (of order $e\, Q^{-1}$) involves seagull and in-flight
contributions associated with one-pion exchange, and the N$^2$LO term (of order $e\, Q^0$)
represents the $(Q/m_N)^2$ relativistic correction to the LO one-body current ($m_N$ denotes the
nucleon mass).  Explicit expressions for all these are listed in Refs.~\cite{Pastore08,Pastore09}.

\begin{figure*}[!htb]
\centering
\includegraphics[width=2\columnwidth]{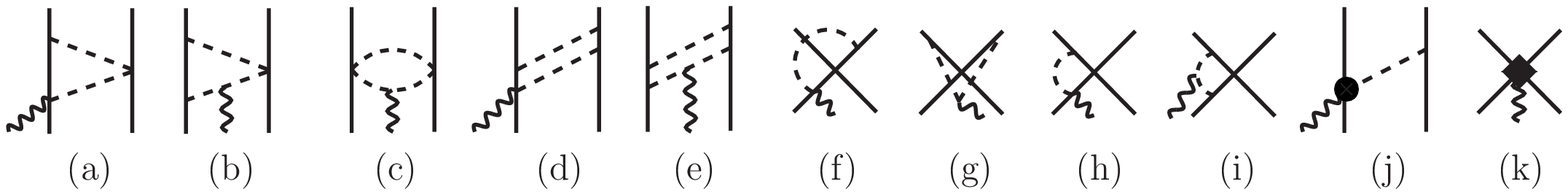}
\caption{Diagrams illustrating two-body currents at N$^3$LO.  Nucleons,
pions, and photons are denoted by solid, dashed, and wavy lines, respectively.
Only one among the possible time orderings is shown for diagrams (a)-(j).}
\label{fig:f1}
\end{figure*}

At N$^3$LO ($e\, Q$) we distinguish three classes of terms~\cite{Pastore09}: i) two-pion exchange currents
at one loop, illustrated by diagrams (a)-(i) in Fig.~\ref{fig:f1}, ii) a tree-level one-pion exchange current
involving the standard $\pi NN$ vertex on one nucleon, and a $\gamma \pi NN$ vertex of order $e\, Q^2$ on the
other nucleon, illustrated by diagram (j), and iii) currents generated by minimal substitution
in the four-nucleon contact interactions involving two gradients of the nucleons' fields as well as by non-minimal
couplings, collectively represented by diagram (k).  A fourth class consisting of $(Q/m_N)^2$ relativistic
corrections to the NLO currents is neglected.

The two-body magnetic moment ($M1$) operator associated with these currents
is conveniently separated into a term dependent on the center-of-mass
position ${\bf R}$ of the two nucleons and one independent of it~\cite{Sachs48}.
The former, known as the Sachs' contribution, is uniquely determined,
via the continuity equation, by the $\chi$EFT potential at order $Q^2$, and
reads~\cite{Pastore09}:
\begin{eqnarray}
{\bm \mu}^{\rm N^3LO}_{\rm Sachs}&=& -\frac{i}{2}\,
e\, ({\bm \tau}_1\times {\bm \tau}_2)_z
\,{\bf R}\times \nabla_k\, v_0^{2\pi}(k) +
\frac{e}{4}\frac{\tau_{1,z}-\tau_{2,z}}{2} 
{\bf R}\nonumber \\
&\times& \Big[ 2\, (C_2 +C_4\, {\bm \sigma}_1\cdot {\bm \sigma}_2)\,
{\bf K} -i\, C_5\, \frac{{\bm \sigma}_1+{\bm \sigma}_2}{2}\times {\bf k} \nonumber \\
&+&C_7\, ({\bm \sigma}_1\,{\bm \sigma}_2\cdot {\bf K}
+{\bm \sigma}_1\cdot{\bf K}\,\,{\bm \sigma}_2) \Big] 
\label{eq:msachs}
\end{eqnarray}
where $v_0^{2\pi}(k)$ is the isospin-dependent part of the two-pion-exchange
chiral potential at order $Q^2$, and $C_2$, $C_4$, $C_5$, and $C_7$ are
low-energy constants (LEC's) entering the contact potential at order $Q^2$.
The function $v_0^{2\pi}(k)$ is defined as
\begin{eqnarray}
v_0^{2\pi}(k)&=&\frac{1}{48 \pi^2\,F_{\pi}^4}\,
G(k)\, \Big[4\, m_{\pi}^2\, (1+4\, g_A^2-5\, g_A^4) \nonumber \\
&+&k^2\, (1+10\, g_A^2 - 23\, g_A^4)-\frac{48\, g_A^4 m^4_\pi}
{4\,  m^2_\pi+k^2}\Big] \ .
\end{eqnarray}

The translationally invariant $M1$ operators associated with pion
loops [diagrams (a)-(i) in Fig.~\ref{fig:f1}], the one-pion-exchange current of order $e\, Q$ [diagram (j)],
and contact currents due to non-minimal couplings [diagram (k)] are given, respectively, by~\cite{Pastore09}
\begin{eqnarray}
{\bm \mu}^{\rm N^3LO}_{\rm loop}&=&\frac{e\,g_A^2}{8\,\pi^2F_{\pi}^4}\,\tau_{2,z}
\Big[F_0(k) \,{\bm \sigma}_1 - F_2(k)\, \frac{{\bf k}\,{\bm \sigma}_1\cdot{\bf k}}{k^2}\Big] \nonumber \\
&+& \frac{e\,g_A^2}{2\,\pi^2F_{\pi}^2}\,\tau_{2,z}\,
\left(C_S\, {\bm \sigma}_2-C_T\,{\bm \sigma}_1\right) + 1 \rightleftharpoons 2 
\label{eq:mloop} \\
{\bm \mu}^{\rm N^3LO}_{\rm tree}&=& e \frac{g_A}{F_\pi^2}
\Big[ \left(d_8^{\, \prime}\, \tau_{2,z} + d_9^{\, \prime}\, {\bm \tau}_1 \cdot {\bm \tau}_2 \right){\bf k}\nonumber \\
\!\!&-&\!\!d_{21}^{\, \prime}\, ({\bm \tau}_1\times{\bm \tau}_2)_z 
{\bm \sigma}_1\times {\bf k}\Big] 
\frac{{\bm \sigma}_2\cdot {\bf k}}{k^2+m_\pi^2}+ 1 \rightleftharpoons 2 
\label{eq:mtree}\\
{\bm \mu}^{\rm N^3LO}_{\rm CT}&=&-e\, C_{15}^\prime \,{\bm \sigma}_1 
\!-\!e\, C_{16}^\prime (\tau_{1,z}-\tau_{2,z})\,
 {\bm \sigma}_1 \!+\! 1 \rightleftharpoons 2  
\label{eq:mct}
\end{eqnarray}
where $d_8^{\, \prime}$, $d_9^{\, \prime}$,
$d_{21}^{\, \prime}$,  $C_{15}^\prime$, and $C_{16}^\prime$, 
are additional LEC's to be determined as discussed below, while
the functions $F_i(k)$ are defined as
\begin{eqnarray}
 F_0(k)&=&1 -2\, g_A^2+\frac{  8\,g_A^2\, m_\pi^2 }{k^2+4\, m_\pi^2}
+G(k)\,\Bigg[ 2-2\, g_A^2\nonumber \\
&-&\frac{  4\,(1+g_A^2)\, m_\pi^2 }{k^2+4\, m_\pi^2}
+\frac{16\, g_A^2 \,m_\pi^4 }{(k^2+4\, m_\pi^2)^2} \Bigg] 
\label{eq:f0k} \\
 F_2(k)&=&2-6\, g_A^2+ \frac{  8\,g_A^2\, m_\pi^2 }{k^2+4\, m_\pi^2}
+G(k)\, \Bigg[4\, g_A^2\nonumber \\
&-&\frac{  4\,(1+3\, g_A^2)\, m_\pi^2 }{k^2+4\, m_\pi^2}
+\frac{16\, g_A^2 \,m_\pi^4 }{(k^2+4\, m_\pi^2)^2} \Bigg] \ .
\label{eq:f2k}
\end{eqnarray}
It is interesting to note that the constant $2-6\, g_A^2$ in $F_2(k)$
would lead to a long-range contribution of the type
\begin{displaymath}
\left[ \tau_{2,z}\, ({\bm \sigma}_1\cdot {\bm \nabla}){\bm \nabla}
+1 \rightleftharpoons 2 \right]1/r
\end{displaymath}
in the magnetic moment, which is, however, fictitious in the
present context of an effective field theory valid at low momenta, since in
performing the Fourier transform the high momentum components
are suppressed by the cutoff $\overline{C}_\Lambda(k)$ (see below).

The isovector part of ${\bm \mu}^{\rm N^3LO}_{\rm tree}$ has
the same structure as the $M1$ operator
involving $N$-$\Delta$ excitation~\cite{Pastore08}, to which
it reduces if the following identifications are made:
$d_{21}^{\, \prime}/d_8^{\, \prime}=1/4$, and
$d_8^{\, \prime} = 4\, \mu^* h_A /(9\, m_N \,\Delta)$,
where $h_A$ is the $\pi N \Delta$ coupling constant, $\mu^*$
is the $N\Delta$-transition magnetic moment, and $\Delta$ is
the $\Delta$-$N$ mass difference, $\Delta=m_\Delta-m_N$.
In this resonance saturation picture, the term proportional to
$d_8^{\, \prime}$ can also be interpreted as due to the $\omega\pi\gamma$ transition current,
ignoring $\omega$-meson propagation
(see Ref.~\cite{Carlson98} and references therein), in which case
$d_8^{\, \prime}=g_{\omega\pi\gamma}g_{\omega NN} F_\pi/m_\omega^3$,
where $g_{\omega\pi\gamma}$ is the $\omega\pi\gamma$ transition coupling constant,
$g_{\omega NN}$ is the $\omega NN$ vector coupling constant, and $m_\omega$ is the
$\omega$-meson mass.  Similarly, the isoscalar part of 
${\bm \mu}^{\rm N^3LO}_{\rm tree}$ reduces to the $\rho\pi\gamma$ $M1$ operator,  if $d_9^{\, \prime}=g_{\rho\pi\gamma}g_{\rho NN} F_\pi/m_\rho^3$,
where $g_{\rho\pi\gamma}$ is the $\rho\pi\gamma$ transition coupling constant,
$g_{\rho NN}$ is the $\rho NN$ vector coupling constant,
and $m_\rho$ is the $\rho$-meson mass.

Currents in $\chi$EFT at N$^3$LO have also been derived, using different
formalisms, by Park {\it et al.} in Ref.~\cite{Park96} and, more recently,
by K\"olling {\it et al.} in Ref.~\cite{Koelling09}.  The derivation in
Ref.~\cite{Park96} is based on covariant perturbation theory, and includes
only the contribution of irreducible diagrams.  Consequently, the cancellations
occurring between the latter and recoil-corrected diagrams
are lacking, and the resulting ${\bm \mu}^{\rm N^3LO}_{\rm loop}$ (in
particular, its isospin structure) is different from that
given here.  In addition, the authors of Ref.~\cite{Park96}
neglect the terms in ${\bm \mu}^{\rm N^3LO}_{\rm loop}$ proportional
to the LEC's $C_S$ and $C_T$ as well as those in ${\bm \mu}^{\rm N^3LO}_{\rm Sachs}$
proportional to the $C_i$, $i$=2, 4, 5, and 7.

The derivation in Ref.~\cite{Koelling09} uses time-ordered perturbation
theory in combination with a unitary transformation that decouples, in the
Hilbert space of nucleons and pions, the states consisting of nucleons only
from those containing, in addition, pions~\cite{Epelbaum98}.  The resulting
expressions for the two-pion-exchange currents, the only ones considered
by the authors of Ref.~\cite{Koelling09}, are in agreement with those
obtained in Ref.~\cite{Pastore09}.

\section{$M1$ observables in $A$=2--4 systems}
\label{sec:res}

In the present contribution we report on the first stage of a research
program aimed at studying electromagnetic observables of
light nuclei, and particularly radiative capture processes in
the three- and four-nucleon systems, within a consistent
$\chi$EFT framework, {\it i.e.}~with the one-loop potential and
currents discussed in the previous sections.  Here, we
present results for $M1$ transitions in $A$=2--4 nuclei
obtained in the hybrid approach, i.e., by evaluating the
matrix elements of the $\chi$EFT $M1$ operators between wave
functions obtained from realistic potentials.  We consider the Argonne
$v_{18}$~\cite{Wiringa95} and chiral N$^3$LO~\cite{Entem02}
two-nucleon potentials in combination with the Urbana-IX~\cite{Pudliner97}
and chiral~\cite{Gazit09} (N2LO) three-nucleon potentials.
These models, denoted as AV18/UIX and N3LO/N2LO,  provide an excellent description
of three- and four-nucleon bound and scattering state properties, including
binding energies, radii, and effective range expansions~\cite{Kievsky08}.
The AV18/UIX model has also been used in a recent (hybrid)
calculation of the astrophysical factor for
the $p$-$p$ and $p$-$^3$He fusion reactions by weak capture at the keV energies
relevant in the interior of the Sun ~\cite{Park03}.

Neutron and proton radiative captures on $^2$H, $^3$H and $^3$He are
particularly challenging from the standpoint of nuclear few-body
theory.  This can be appreciated by comparing the measured values
for the cross sections of thermal neutron radiative capture on
$^1$H, $^2$H, $^3$He.  Their respective values in mb are: ($332.6\pm 0.7$)~\cite{Mughabghab81},
($0.508 \pm 0.015$)~\cite{Jurney82}, and ($0.055\pm 0.003$)~\cite{Wolfs89}.
Thus, in going from $A$=2 to 4 the cross section has dropped by
almost four orders of magnitude.  These processes are induced by
$M1$ transitions between the initial two-cluster state
in relative S-wave and the final bound state.  The $^3$H and $^4$He
wave functions, respectively $\Psi_3$ and $\Psi_4$, are approximately
eigenfunctions of the one-body $M1$ operator ${\bm \mu}$, namely
$\mu_z  \Psi_3 \simeq \mu_p \Psi_3$ and $\mu_z  \Psi_4 \simeq 0$,
where $\mu_p$=2.793 n.m.~is the proton magnetic moment---the experimental
value of the $^3$H magnetic moment is 2.979 n.m, while $^4$He has no
magnetic moment.  These relations would be exact, if the $^3$H
and $^4$He wave functions were to consist of the symmetric S-wave
term only.  In fact, tensor components in the nuclear potentials
generate significant D-state admixtures, that partially spoil
this eigenstate property.  To the extent that it is approximately
satisfied, though, the matrix elements $\langle\Psi_3\!\mid\!\mu_z\!\mid\!\Psi_{1+2}\rangle$
and $\langle\Psi_4\!\mid\!\mu_z\!\mid\!\Psi_{1+3}\rangle$ vanish
due to orthogonality between the initial and final states.
This orthogonality argument fails in the case of the deuteron,
since then $\mu_z\Psi_2\simeq (\mu_p-\mu_n) \,\phi_2(S)\, \chi^0_0\, \eta^1_0\,$,
where $\chi^S_{M_S}$ and $\eta^T_{M_T}$ are two-nucleon spin and isospin
states, respectively.  The $M1$ operator can therefore
connect the large S-wave component $\phi_2({\rm S})$ of the deuteron
to a $T$=1 $^1$S$_0$ $n$-$p$ state (the orthogonality between the latter
and the deuteron follows from the orthogonality between their respective
spin-isospin states).

As a result of this suppression, the $n$-$d$, $p$-$d$, $n$-$^3$He, and
$p$-$^3$H radiative (as well as $p$-$^3$He weak) captures are very
sensitive to small components in the wave functions, particularly
the D-state admixtures generated by tensor forces, and to many-body
terms in the electromagnetic (and weak) current operators.

\begin{figure}[bthp]
\includegraphics[width=3.5in]{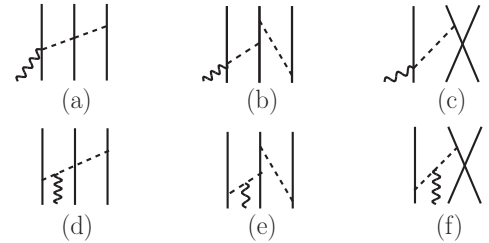}
\caption{Diagrams illustrating three-body currents at N$^3$LO.  Notation
as in Fig.~\protect\ref{fig:f1}.  Their contribution vanishes, see text for discussion.}
\label{fig:f2}
\end{figure}

The $\chi$EFT $M1$ operators discussed
in the previous section, regularized by a
cutoff $\overline{C}_\Lambda(k)={\rm exp}(-k^4/\Lambda^4)$ with $\Lambda$
in the range between 500 MeV and 700 MeV, have been used~\cite{Girlanda09}
to study the magnetic moments of the deuteron and trinucleons,
and the $np$, $nd$, and $n\,^3$He radiative captures  at thermal neutron
energies.  At N$^3$LO there are no three-body currents, since
the contributions of diagrams (a) and (d) in Fig.~\ref{fig:f2}, involving the $\pi\pi NN$
vertex, vanish, while those due to the irreducible and recoil-corrected reducible
diagrams---only irreducible diagrams are shown in panels (b)-(c) and (e)-(f) of
Fig.~\ref{fig:f2}---exactly cancel out.

\begin{figure*}[!htb]
\centering
\includegraphics[width=1.25\columnwidth]{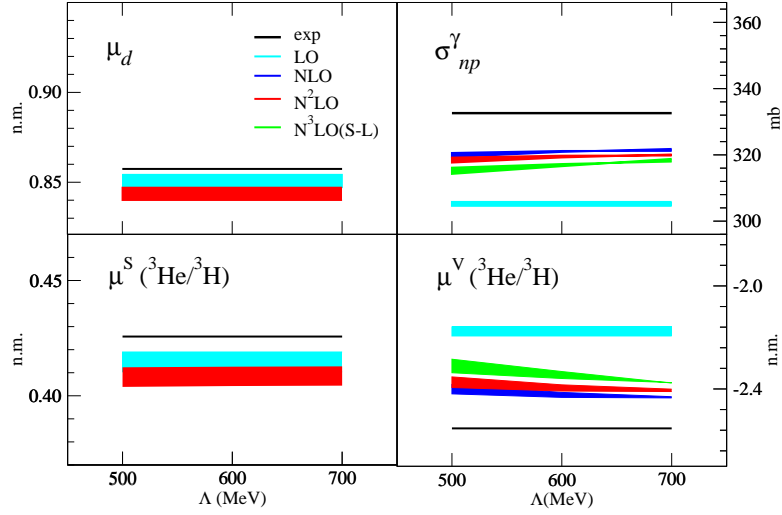}
\vspace{0.5cm}
\caption{Results for the deuteron and trinucleon
isoscalar and isovector magnetic moments, and $np$ radiative
capture, obtained by including cumulatively the LO, NLO,
N$^2$LO, and N$^3$LO(S-L) contributions.  See text for discussion.}
\label{fig:f3}
\end{figure*}

We now turn our attention to the determination of the LEC's $d_8^{\, \prime}$,
$d_9^{\, \prime}$, $d_{21}^{\, \prime}$,  $C_{15}^\prime$, and $C_{16}^\prime$.
In principle, the $d_i^{\, \prime}$ could be fitted to pion photoproduction
data on a single nucleon, or related to hadronic coupling constants
(although $g_{\omega NN}$ and $g_{\rho NN}$
are rather poorly known) by resonance saturation arguments.
Indeed, this latter strategy is used in a series of calculations, based on the $M1$
operators derived in Ref.~\cite{Park96}, of the $np$, $nd$, and $n\, ^3$He radiative
captures, and magnetic moments of $A$=2 and 3 nuclei~\cite{Park96,Park00}.
Here, however, we assume $d_{21}^{\, \prime}/d_8^{\, \prime}=1/4$ as suggested
by resonance saturation, and rely on nuclear data to constrain the remaining
four LEC's.  The values obtained by reproducing the experimental $np$ cross section
and magnetic moments of the deuteron and trinucleons are listed in Tables~\ref{tb:tab1}
and~\ref{tb:tab1a}.
Note that the adimensional values reported there are in units of powers of $\Lambda$, i.e., we
have defined $d_9^{\, \prime}=d^S_1/ \Lambda^2$, $C_{15}^\prime=d^S_2/ \Lambda^4$,
$d_{21}^{\, \prime}=d^V_1/ \Lambda^2$, and $C_{16}^\prime=d^V_2/\Lambda^4$
and the superscripts $S$ and $V$ denote the isoscalar and isovector content of the
associated operators.
\begin{table}
\caption{Adimensional values of the isoscalar LEC's corresponding to cutoff parameters $\Lambda$
in the range 500--700 MeV obtained for the AV18/UIX (N3LO/N2LO) Hamiltonian.  See text for explanation.}
\label{tb:tab1}
\begin{tabular}{lcc}
\hline\noalign{\smallskip}
$\Lambda$   & $d^S_1\times 10^2$ & $d^S_2$   \\
\noalign{\smallskip}\hline\noalign{\smallskip}
500  & --8.85 (--0.225)  &  --3.18 (--2.38)  \\
600  & --2.90 (9.20)  &   --7.10 (--5.30)     \\
700  &  6.64 (20.4) & --13.2 (--9.83)  \\
\noalign{\smallskip}\hline
\end{tabular}
\end{table}

\begin{table}
\caption{Adimensional values of the isovector LEC's corresponding to cutoff parameters $\Lambda$
in the range 500--700 MeV obtained for the AV18/UIX (N3LO/N2LO) Hamiltonian.  See text for explanation.}
\label{tb:tab1a}
\begin{tabular}{lcc}
\hline\noalign{\smallskip}
$\Lambda$   & $d^V_1$ & $d^V_2$  \\
\noalign{\smallskip}\hline\noalign{\smallskip}
500  & 5.18 (5.82) & --11.3 (--11.4) \\
600  & 6.55 (6.85) & --12.9 (--23.3)  \\
700  &  8.24 (8.27) & --1.70 (--46.2) \\
\noalign{\smallskip}\hline
\end{tabular}
\end{table}

In the discussion to follow, we will refer
to the terms in Eqs.~(\ref{eq:msachs}) and~(\ref{eq:mloop}) as N$^3$LO(S-L)
and to those in Eqs.~(\ref{eq:mtree}) and~(\ref{eq:mct}) as N$^3$LO(LECs).
In Fig.~\ref{fig:f3} we show results obtained by including cumulatively
the contributions at LO, NLO, N$^2$LO, and N$^3$LO(S-L) for the
deuteron ($\mu_d$) and $^3$He/$^3$H isoscalar ($\mu^S$) magnetic
moments (left panels), and for the $np$ radiative capture cross section
($\sigma_{np}^\gamma$) at thermal energies and $^3$He/$^3$H
isovector ($\mu^V$) magnetic moment (right panels).
The NLO and N$^3$LO(S-L) $M1$ operators are purely isovector, and
hence do not contribute to $\mu_d$ and $\mu^S$, while
the Sachs' term in the N$^3$LO(S-L) operator vanishes
in $A$=2 systems.  The band represents the spread in
the calculated values corresponding to the two Hamiltonian models considered
here (AV18/UIX and N3LO/N2LO).  The sensitivity to short-range
mechanisms (effective at internucleon separations less than
$(2\, m_\pi)^{-1}$, say) as encoded in the cutoff $\overline{C}_\Lambda(k)$ and
in the rather different short-range behaviors of the adopted potentials,
remains quite weak for all observables.  Of course, taking into account
the N$^3$LO contribution with the LEC values listed
in Tables~\ref{tb:tab1} and~\ref{tb:tab1a} reproduces the experimental data
represented by the black band (to accommodate errors, although
these are negligible in the present case).
The contributions at LO and NLO have
the same sign, while those at N$^2$LO and N$^3$LO(S-L)
have each opposite sign, and tend to increase the difference
between theory and experiment.

\begin{figure*}[!htb]
\centering
\includegraphics[width=1\columnwidth]{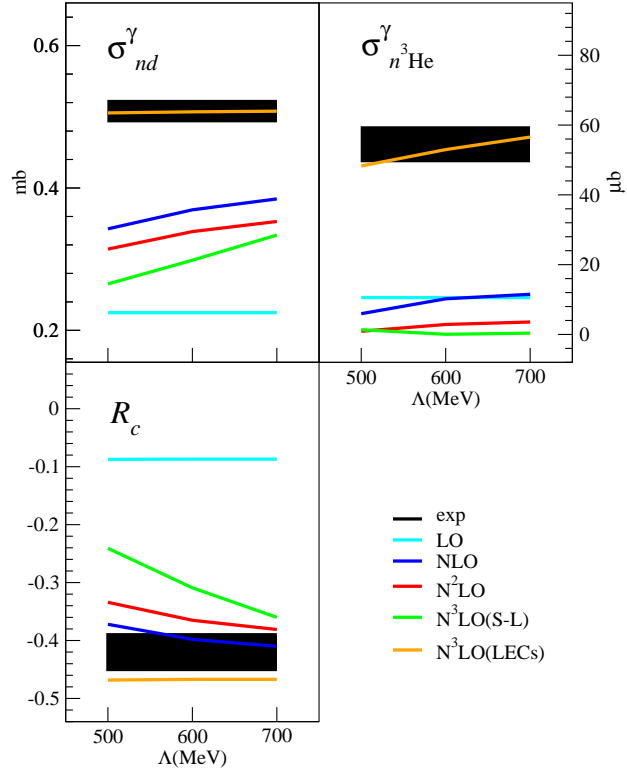}
\vspace{0.5cm}
\caption{Results for $\sigma_{nd}^\gamma$ (left top panel),
$\sigma_{n\, ^3{\rm He}}^\gamma$ (right top panel),
and $R_c$ (left bottom panel), obtained by including cumulatively the LO, NLO,
N$^2$LO, N$^3$LO(S-L), and N$^3$LO(LECs) contributions.
See text for discussion.}
\label{fig:f4}
\end{figure*}

Having fully constrained the $M1$ operator up to N$^3$LO,
we are now in a position to present a preliminary set of predictions, shown in
Fig.~\ref{fig:f4}, for the $nd$ and $n\, ^3$He radiative
capture cross sections, denoted as $\sigma_{nd}^\gamma$
and $\sigma_{n\, ^3{\rm He}}^\gamma$, and the photon
circular polarization parameter $R_c$ resulting from the
capture of polarized neutrons on deuterons.  The experimental
data (black bands) are from Ref.~\cite{Jurney82} for $nd$ and Ref.~\cite{Wolfs89} for $n\, ^3$He.
In this first stage, we have used only the AV18/UIX (N3LO/N2LO) wave functions
for the $A$=3 ($A$=4) processes.

Results obtained with the complete N$^3$LO operator are shown by the orange
lines labeled N$^3$LO(LECs), and are in very satisfactory agreement
with data.  Their sensitivity to the cutoff is negligible for $nd$ and
at the 5\% level for $n\, ^3$He.  As already remarked,
these processes are strongly suppressed at LO:
the calculated $\sigma_{nd}^\gamma$(LO) and $\sigma_{n\, ^3{\rm He}}^\gamma$(LO)
are less than half and a factor of five smaller than the measured values.
In the case of $n\, ^3$He, the matrix element at NLO is
of opposite sign and twice as large (in magnitude) compared to that at LO, hence
$\sigma^\gamma_{n\, ^3{\rm He}}$ at LO and LO+NLO
are about the same, as seen in Fig.~\ref{fig:f4}.  For $nd$, however, the LO and NLO contributions
interfere constructively.  For both $nd$ and $n\, ^3$He, the N$^2$LO and N$^3$LO(S-L) corrections
exhibit the same pattern discussed in connection with Fig.~\ref{fig:f3}.  The N$^3$LO(LECs)
contributions are large and crucially important for bringing theory into agreement with experiment.

Song {\it et al.}~(2009)~\cite{Park00} and Lazauskas {\it et al.}~\cite{Park00}
have reported values for the $nd$ and $n\, ^3$He capture cross sections about 6\% and
15\% smaller than measured, with a significantly larger sensitivity  (estimated at $\simeq 15$\%
for both processes) to the cutoff.  We have already noted the differences
between the N$^3$LO $M1$ operators used by these authors (in particular, their reliance
on resonance saturation to constrain the LEC's entering ${\bm \mu}^{\rm N^3LO}_{\rm tree}$) and
those in the present work.

The work of R.S.\ is supported by the U.S.~Department of Energy,
Office of Nuclear Physics under contract DE-AC05-06OR23177.
Some of  the calculations were made possible by grants of computing
time from the National Energy Research Supercomputer Center.
\end{document}